# The "Long Secondary Period (LSP) Variables": Overview and Some Analysis

**John R. Percy**
**Mayank H. Shenoy**
*Department of Astronomy and Astrophysics, and Dunlap Institute for Astronomy and Astrophysics, University of Toronto, 50 St. George Street, Toronto ON M5S 3H4, Canada; john.percy@utoronto.ca*

**Abstract**   We briefly review the phenomenon of "long secondary periods" (LSPs) in red giants, and the "LSP variable stars" classification introduced in the All-Sky Automated Survey for Supernovae (ASAS-SN) variable star catalog; they are red giant Long Period Variables (LPVs) in which their LSP variability is significantly greater than their pulsational variability. We then describe and discuss the results of a period and amplitude analysis of a random sample of 35 LSP variables in the ASAS-SN catalog, using ASAS-SN data and the AAVSO VStar time-series analysis software. The pulsation period and amplitude, and LSP, all increase with increasing luminosity or size of the star, as expected. The behavior of the LSP amplitude is more complicated; it appears to be larger in moderate-luminosity stars, and smaller in low- and high-luminosity stars. In particular, it is relatively small in a sample of 27 Mira stars, analyzed separately using AAVSO visual data. These results are discussed in the context of the current model for the LSP phenomenon, namely that it is caused by eclipses of the red giant star by a dust-enshrouded companion.

**1. Introduction**

Red giant stars vary in brightness in complex ways, and for a multitude of reasons. The All-Sky Automated Survey for Supernovae variable star catalog (ASAS-SN; Shappee *et al.* 2014; Kochanek *et al.* 2017) has recently introduced a new subclass of red giant variables—ones whose variability is dominated by a "long secondary period" (LSP) which is very common in red giants but, until recently, poorly understood. The purpose of the present paper is to introduce these variables and the LSP phenomenon, to present an analysis of a sample of 35 LSP variables from the ASAS-SN catalog, and to discuss why they continue to be rather enigmatic.

Red giants are unstable to low-order radial pulsation. The periods and amplitudes are greater for larger, more luminous stars, ranging from a few days and a few thousandths of a magnitude in the least luminous stars, to hundreds of days and several magnitudes in the most luminous stars. The smaller-amplitude variables are classified as semiregular (SR) or irregular (L) and sometimes OGLE small amplitude red giants (OSARG); the variables with pulsation amplitudes of 2.5 or greater are classified as Mira (M) stars. Collectively, these stars are often referred to as Long Period Variables (LPVs), but the terminology is complicated; the LPV Observing Section of the American Association of Variable Star Observers (AAVSO) includes yellow supergiant variables such as RV Tauri stars. In the Hertzsprung-Russell Diagram of luminosity vs. surface temperature, these stars are on or approaching the upper end of the hydrogen-burning Red Giant Branch, or the subsequent helium-burning Asymptotic Giant Branch. As the star becomes more luminous, mass loss increases, and eventually becomes so great that the star loses the outer half of its mass, which forms a slowly-expanding planetary nebula around the star. The core of the star becomes a white dwarf.

One-third to one-half of red giants display another form of variability—long secondary periods, 5 to 10 times the radial pulsation period. These LSPs were identified in large numbers by photographic photometry, especially at the Harvard College Observatory (O'Connell 1933; Payne-Gaposchkin 1954; Houk 1963). Houk (1963) listed over a hundred LPVs with LSPs. Subsequent visual observations of Houk's stars confirmed that the vast majority of these LSPs were correct (Percy 2022).

The Harvard photographic program was gradually phased out, but the study of LSPs continued, thanks in part to visual observations by skilled amateur astronomers, working through organizations such as the AAVSO; see Percy and Deibert (2016), for instance, for an example of the analysis of these observations.

By 1980, there were two more important contributors to the study of LSPs: remote robotic observatories (e.g. a survey of two dozen red giants by Percy *et al.* 2001), and long-term photoelectric photometry by skilled amateur astronomers (e.g. through the photoelectric photometry program of the AAVSO: Percy *et al.* 1989), made possible by the availability of reasonably-priced, off-the-shelf photometers.

The 1990s brought a wave of massive automated long-term photometric surveys of large numbers of stars by the professional astronomical community (ASAS: All-Sky Automated Survey, MACHO: MAssive Compact Halo Objects, and OGLE: Optical Gravitational Lensing Experiment). LSPs were so prevalent that they were quickly discovered in red giants in large numbers. Wood (2000), in particular, called attention to the period-luminosity sequences for red giants in the Large Magellanic Cloud. There were sequences for low-order modes of pulsation, and also a well-defined sequence for LSPs.

But what was the cause of the LSPs? Wood *et al.* (2004) referred to LSPs as "the only unexplained type of large-amplitude stellar variability known at this time." Was it due to non-radial pulsation, some form of eclipse, rotation, convection cells, magnetic phenomena, or something else? For two decades, the problem of LSPs was a vigorous area of study.

A proposal by Soszyński *et al.* (2021) seemed to provide an answer: "The LSP light changes are due to the presence of a dusty cloud orbiting the red giant, together with a companion, and obscuring the star once per orbit.... In this scenario, the low-mass companion is a former planet that has accreted a significant amount of mass from the envelope of its host star and grown into a brown dwarf."

This proposal was based on a study of about 700 LPVs in the OGLE survey (Udalski *et al.* 2015) of the Magellanic Clouds and Galactic bulge, using both visual and infrared photometry. The strongest new evidence for this proposal was the observation that, in the infra-red, there is a secondary eclipse which occurs when the very cool companion and its cloud—which are infra-red sources but not optical sources—are eclipsed by the red giant star.

There are, nevertheless, some mysteries arising from this proposal. Surveys show that about one-third of red giants show LSPs. But the Soszyński *et al.* (2021) mechanism implies that there would also be systems in which the companion's orbit was seen face-on, in which case there would be no eclipse, and no LSP, so the actual fraction of systems having dusty companions would be even greater than the fraction showing LSPs. It is surprising that such systems had not been discovered much earlier, among the brighter, nearby, well-studied red giants.

LSPs appear to be common in globular clusters (Lebzelter and Wood 2005; Percy and Gupta 2021; Kim and Percy 2022), as well as in the Magellanic Clouds. That would imply that, in order to provide the mechanism for LSPs, both planet formation and dust formation were common enough in these ancient star systems with low abundances of elements heavier than helium to provide the ingredients for the LSP process.

Another mystery is the relatively tight LSP-luminosity relation found by Wood (2000). Soszyński *et al.* (2021) deliberately selected stars for their study which lay in the center of the LSP-luminosity relation to ensure that they were truly stars with LSPs so in their study there is a bias or selection effect in producing a tight period-luminosity relation. This bias is not present in Wood's (2000) study.

However, the luminosity of the red giant depends on its radius. The LSP depends on the radius of the companion's orbit. Why should these be closely related? Kim and Percy (2022) discussed this, and estimated that, for the stars in their sample, the latter should be about twice the former. This seems reasonable: the orbit of the companion cannot be too close to the star, and if it is too far from the star, the accretion process may be inefficient, and/or the probability of eclipses may also be smaller, since the orbit needs to be seen edge-on in order to produce LSP variations.

The present paper deals with another mystery: there are red giant stars in which the pulsation period and amplitude are small, suggesting that the luminosity of the red giant is low, yet an LSP is present with an amplitude larger than that of the pulsation. If the luminosity of the red giant is low, its mass-loss rate will presumably be low. In that case, how did the companion accrete enough material to become massive and dusty enough to cause the LSP phenomenon? In the ASAS-SN variable star catalog, there are 185 such stars in which the LSP variation dominates the pulsation variation. The catalog classifies these stars as "LSP variables," though that classification is not (yet) found in the *General Catalogue of Variable Stars* (Samus *et al.* 2017). In the present paper, we analyze a sample of 35 such stars, and study the periods and amplitudes of the LSP and the pulsation in order to study the systematics of this new "class" of stars.

Because the results of this first project suggested that more luminous LPVs had smaller LSP amplitudes, we also analyzed visual measurements of a sample of 27 Mira stars—larger, more luminous variables with periods of hundreds of days, and ranges of 2.5 magnitudes or more.

Pawlak (2023), using a different approach, analyzed OGLE observations of 1663 Mira stars in the LMC, and concluded that seven percent of the Mira stars in the sample might have LSPs. This is a much smaller percentage than found among SR variables in similar surveys.

## 2. Data and analysis

From the ASAS-SN variable star website and catalog (Shappee *et al.* 2014; Jayasinghe *et al.* 2018, 2019), data on a random sample of 35 stars classified as LSP variables were downloaded and analyzed with careful light-curve analysis and time-series analysis using the AAVSO VSTAR software package (Benn 2013). The sample size was determined by the fact that detailed star-by-star analysis could be accomplished within the time available for this student project. It was large enough to point to any interesting and/or unusual results, which could be followed up, such as in our parallel study (Percy and Zhitkova 2023).

The stars and results are listed in Table 1, which gives the ASAS-SN star name, the apparent V magnitude and absolute K magnitude, and the period and (semi) amplitude of the LSP (LSP, LSP-A) and of the dominant pulsation period (PP, PP-A). The absolute K magnitude $M_K$, which is more representative of the luminosity of red variables than the absolute V magnitude, was determined from the K magnitude, Gaia distance, and interstellar reddening, as given in the ASAS-SN variable star catalog. Figures 1 and 2 show ASAS-SN light curves of two representative LSP variables.

Note that the ASAS-SN datasets are only about 1500 days long. This limits the precision of the periods determined from them, especially the LSP, which is typically hundreds of days. The ASAS-SN data also have the usual seasonal gaps in the data (see Figures 1 and 2) which introduce the possibility of spurious "alias" periods in the Fourier spectra.

## 3. Results

The PP increases with increasing luminosity MK, as expected (Figure 3). Much of the scatter is probably due to the fact that the stars may be pulsating in different low-order modes, or possibly a mixture of modes. The pulsation of LPVs is semi-regular at best, in part because they vary significantly in amplitude.

The LSP also increases with increasing luminosity or size of the star (Figure 4), as expected from the Soszyński *et al.* (2021) mechanism. Much of the scatter is due to the difficulty in determining the LSP accurately from the short ASAS-SN datasets. The shortness of the datasets also produces a slight bias in favor of shorter LSPs.

The ratio LSP/PP averages about 9 for all $_{MK}$, but with considerable scatter, which is not surprising, given the difficulty in measuring the LSP accurately from short datasets. This ratio suggests that most of these lower-luminosity red giants are pulsating in the first overtone, as previous results have suggested (e.g. Wood 2000). There are some short-period stars with LSP/PP ratios of 13, which may be second-overtone pulsators, and a few longer-period stars with ratios less than 7, which may be fundamental-mode pulsators.

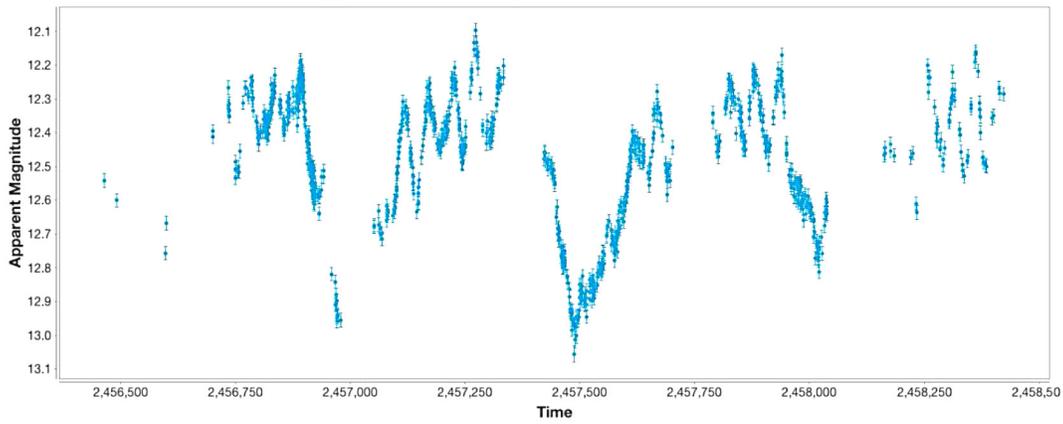

Figure 1. ASAS-SN V light curve of the LSP star ASAS-SN-V J191616.35+475823.7, showing both the LSP variation and the smaller pulsational variation.

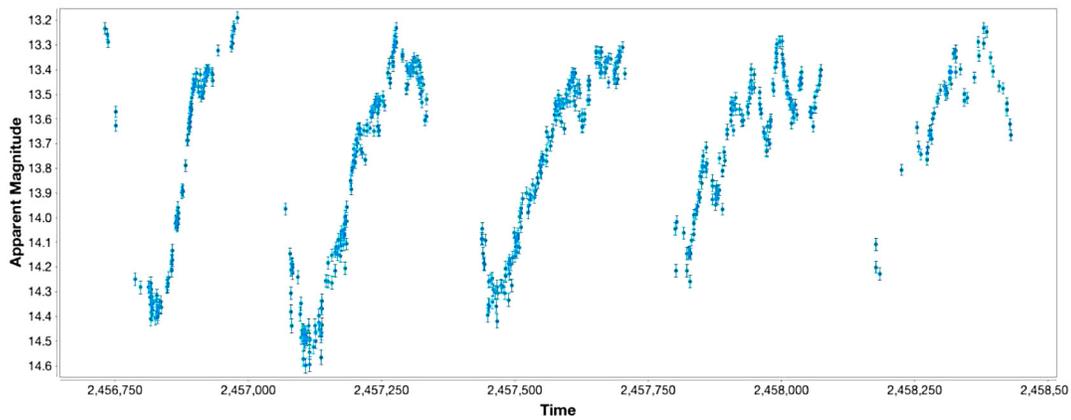

Figure 2. ASAS-SN V light curve of the LSP star ASAS-SN-V J195427.42+474921.2, showing both the LSP variation and the smaller pulsational variation.

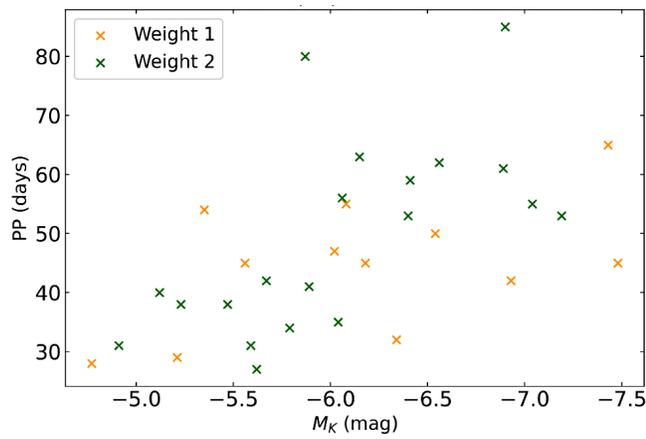

Figure 3. The period-luminosity ($M_K$) relation for the stars in Table 1. The scatter suggests that the stars are not all pulsating in the same mode and/or that the choice of "LSP stars" from the ASAS-SN catalog introduces a selection effect in the sample.

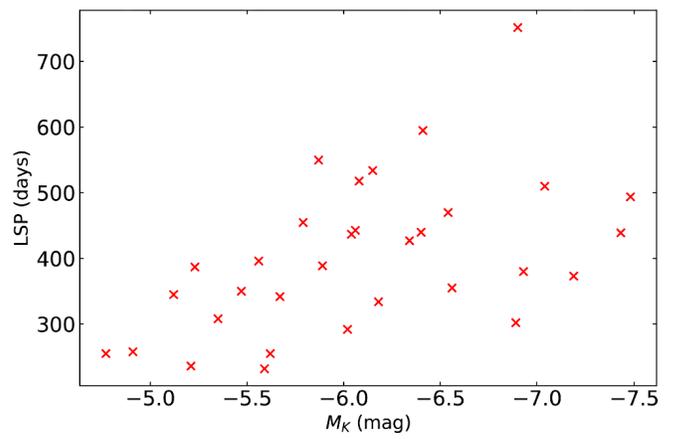

Figure 4. The relation between the LSP and the luminosity $M_K$. There is a general trend, as we would expect from the Soszyński *et al.* mechanism. See text for discussion.

Table 1. Period and amplitude analysis of ASAS-SN observations of LSP stars.

| ASAS-SN-V Name | V | MK | LSP (d) | SA (LSP) | PP (d) | SA (P) |
|---|---|---|---|---|---|---|
| J003836.62+510342.1 | 13.70 | –6.41 | 595 | 0.19 | 59 | 0.120 |
| J020359.53+141132.4 | 12.00 | –5.56 | 396 | 0.39 | 45 | 0.087 |
| J031644.30+790013.7 | 12.49 | –6.06 | 443 | 0.28 | 56 | 0.084 |
| J051411.00+475413.4 | 13.57 | –7.37 | 599 | 0.31 | 30 | 0.110 |
| J100459.29+470246.9 | 12.70 | –5.23 | 387 | 0.08 | 38 | 0.027 |
| J122246.23+152052.5 | 12.01 | –5.12 | 345 | 0.04 | 40 | 0.022 |
| J134521.91+010137.6 | 13.66 | –4.74 | 327 | 0.22 | 14 | 0.055 |
| J134533.58+010408.8 | 12.27 | –5.35 | 308 | 0.10 | 54 | 0.045 |
| J142127.71+463131.7 | 13.20 | –5.47 | 350 | 0.07 | 38 | 0.026 |
| J154021.56+290510.8 | 13.53 | –6.93 | 380 | 0.08 | 42 | 0.027 |
| J162106.89+304132.5 | 13.78 | –4.91 | 258 | 0.07 | 31 | 0.030 |
| J162708.92+261624.2 | 12.61 | –4.77 | 255 | 0.07 | 28 | 0.019 |
| J164815.57–195122.3 | 13.11 | –5.79 | 455 | 0.14 | 34 | 0.041 |
| J173239.14–171835.7 | 12.96 | –6.40 | 440 | 0.25 | 53 | 0.100 |
| J185255.04+290949.6 | 13.90 | –6.35 | 811 | 0.08 | 48 | 0.042 |
| J185735.62–282447.4 | 12.76 | –6.04 | 437 | 0.12 | 35 | 0.048 |
| J185802.67+421553.1 | 13.21 | –6.54 | 470 | 0.08 | 50 | 0.063 |
| J185957.97+404745.5 | 13.48 | –7.19 | 373 | 0.23 | 53 | 0.064 |
| J190119.21+460405.7 | 12.90 | –5.59 | 232 | 0.10 | 31 | 0.035 |
| J190424.97+382941.9 | 13.89 | –6.34 | 427 | 0.30 | 32 | 0.087 |
| J190445.35+320054.7 | 13.02 | –6.15 | 534 | 0.23 | 63 | 0.160 |
| J191456.72+421950.0 | 13.90 | –6.90 | 752 | 0.16 | 85 | 0.110 |
| J191516.97+475851.2 | 12.91 | –7.43 | 439 | 0.11 | 65 | 0.081 |
| J191616.35+475823.7 | 12.47 | –7.04 | 510 | 0.21 | 55 | 0.079 |
| J192201.47+442631.0 | 12.59 | –5.21 | 236 | 0.08 | 29 | 0.024 |
| J192439.93+453046.1 | 13.78 | –7.48 | 494 | 0.12 | 45 | 0.060 |
| J193000.06+413632.3 | 12.57 | –6.02 | 292 | 0.19 | 47 | 0.074 |
| J193220.91+490643.2 | 12.26 | –6.18 | 334 | 0.16 | 45 | 0.023 |
| J193701.71+444317.8 | 12.99 | –6.56 | 355 | 0.08 | 62 | 0.072 |
| J195427.42+474921.2 | 13.76 | –5.67 | 342 | 0.44 | 42 | 0.074 |
| J204505.70+002505.6 | 11.80 | –5.89 | 389 | 0.09 | 41 | 0.041 |
| J212114.88+455019.1 | 13.30 | –6.89 | 302 | 0.08 | 61 | 0.065 |
| J213355.20–001944.3 | 11.50 | –5.62 | 255 | 0.09 | 27 | 0.031 |
| J213751.82+423137.1 | 13.03 | –6.08 | 518 | 0.16 | 55 | 0.050 |
| J233235.02+480155.4 | 13.70 | –5.87 | 550 | 0.15 | 80 | 0.055 |

Table 2. Amplitudes of the LSP variability in some Mira stars.

| Star | PP (days) | LSP-A |
|---|---|---|
| R And | 409 | ≤ 0.25 |
| W And | 396 | ≤ 0.20 |
| R Aur | 458 | ≤ 0.20: |
| T Cam | 373 | ≤ 0.10 |
| T Cas | 445 | ≤ 0.10 |
| o Cet | 332 | 0.53 |
| U Cet | 235 | ≤ 0.20 |
| R CMi | 338 | ≤ 0.10 |
| S CMi | 333 | 0.21 |
| R Cnc | 362 | ≤ 0.12 |
| V Cnc | 272 | ≤ 0.15 |
| S CrB | 361 | 0.23 |
| R Gem | 370 | ≤ 0.20 |
| S Hya | 257 | ≤ 0.15 |
| T Hya | 289 | ≤ 0.15 |
| R Leo | 310 | ≤ 0.15 |
| R Lep | 427 | ≤ 0.10: |
| R LMi | 372 | ≤ 0.20 |
| R Lyn | 365 | ≤ 0.10 |
| V Mon | 340 | ≤ 0.20 |
| U Ori | 369 | 0.20: |
| RZ Per | 369 | ≤ 0.10 |
| Z Sco | 363 | ≤ 0.10 |
| R Tri | 267 | ≤ 0.15 |
| R UMa | 302 | ≤ 0.15 |
| S Vir | 367 | ≤ 0.15 |
| SS Vir | 364 | ≤ 0.10 |

The PP amplitude increases with luminosity (Figure 5), as would be expected from previous results. Almost all the stars have PP amplitudes less than 0.1, and about half have PP amplitudes less than 0.05. This is in part due to the selection effect in choosing "LSP variables," in which, by definition, the LSP amplitude is significantly larger than the PP amplitude.

There is a clear correlation between LSP amplitude and PP amplitude, with the former increasing from 0.05 to about 0.25 when the latter increases from 0.02 to 0.10 (Figure 6). This is to be expected, since both are correlated with luminosity. This reminds us, however, that we are dealing with small and variable amplitudes, so both periods and amplitudes are challenging to determine.

Figure 7 is probably the most interesting in this study, since there have been few, if any, studies of the amplitudes and phase curves of the LSP phenomenon, which is an eclipse phenomenon; Derekas *et al.* (2006) is an exception. The figure suggests that the LSP amplitude is smaller for lower- and higher-luminosity stars, and larger for moderate-luminosity ones.

In part to test this result, we studied a small sample of 27 Mira stars—the largest, highest-luminosity red giants—for LSPs and LSP amplitudes, using visual data from the AAVSO International Database, and VStar. In almost every case, there was no obvious LSP which rose above the noise level, so the best we could do was to give an upper limit to the LSP amplitude, namely, the noise level. The results are given in Table 2 which, in most cases, gives these upper limits. Almost all of these upper limits are in the range 0.10 to 0.20, consistent with the results in Figure 7.

### 4. Discussion

In the course of a parallel study—inspired by this one—of the variability properties of LPVs in the AAVSO Binocular Program (Percy and Zhitkova 2023), we identified stars which had catalog periods of several hundred days, whereas pulsation periods 5 to 10 times shorter could be seen in light curves from ASAS-SN or AAVSO photometry. These include RW Boo, BM Eri, PV Peg, CI Phe, RY UMa, VW UMa, and GO Vel. These stars are therefore "LSP variables," and their catalog periods are LSPs. Catalog users should be aware that the catalog periods of LPVs are not always the pulsation periods.

The ratios of LSP to PP are consistent with previous results, and with the idea that lower-luminosity red giants are more likely to pulsate in the first overtone. In this sense, our sample is similar to other samples of pulsating red giants.

Figure 7 is perhaps the most interesting result of this study, though it is far from definitive. The LSP amplitude is a measure of the eclipse coverage by the dust-enshrouded companion. This presumably depends on the relative sizes of the star and companion (or, more precisely, the effective size of its dusty envelope), and on the inclination of the companion's orbit. If the size of the companion was independent of luminosity, we would expect the LSP amplitude to decrease with increasing luminosity and size of the star, which it does. But the LSP amplitude also becomes smaller in lower-luminosity, smaller stars. This suggests that the effective size of the companion and its envelope is *smaller* in smaller, lower-luminosity stars. This could be because such stars have weaker winds, and have not been able to transfer as much dust to their companion.

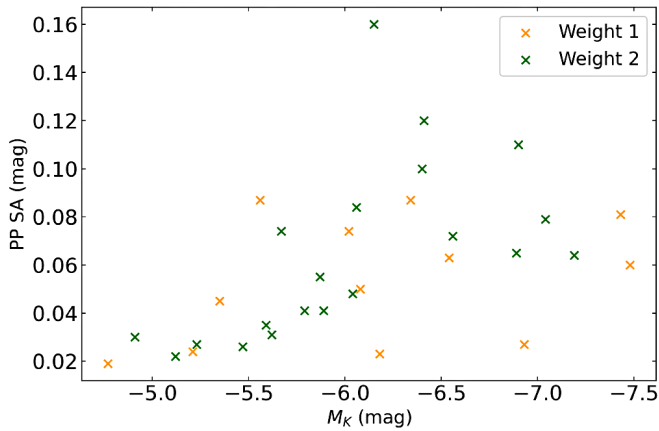

Figure 5. The relation between the pulsation amplitude and the luminosity $M_K$. There is a positive trend, as expected, but the choice of "LSP variables" from the ASAS-SN catalog undoubtedly introduces a bias in favor of small pulsation amplitudes.

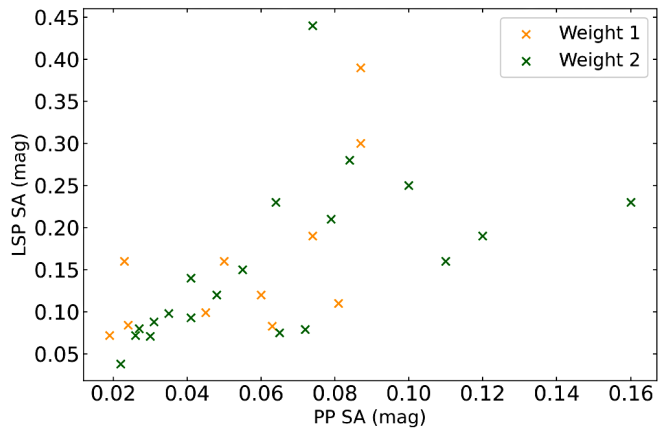

Figure 6. The relation between LSP amplitude and pulsation amplitude. There is a distinct positive trend, as would be expected, since both are positively correlated with the luminosity.

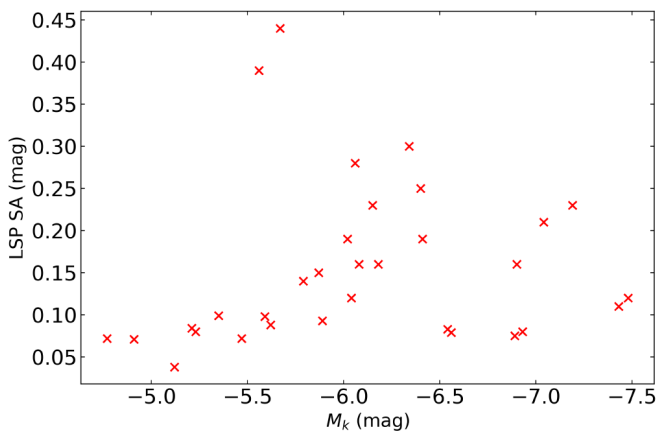

Figure 7. The relation between LSP amplitude and luminosity $M_K$. The LSP amplitude is smaller for low- and high-luminosity stars, especially when the results for Mira stars are included.

Derekas *et al.* (2006) measured LSP amplitudes in a much larger sample of stars, and found the same pattern as we did, namely that the distribution of LSP amplitudes peaked in moderate-luminosity stars and, at every luminosity, there was a range of amplitudes. However, in those authors' opinion, the results "...argue for pulsation rather than binarity as the cause of the LSP phenomenon."

It would obviously be desirable to have more information on the relationship between the LSP amplitude and the luminosity of the star. That was the motivation for studying the LSPs in the sample of Mira stars in Table 2. We also examined LSP amplitudes in two overlapping samples: 37 LPVs with LSPs (Percy 2022) and listed in Houk (1963), and several dozen red variables in the AAVSO Binocular Program (Percy and Zhitkova 2023). For stars with short LSPs (400–800 days), and therefore assumed to be smaller stars, the LSP amplitudes were less than 0.15 and averaged less than 0.10. For stars with moderate LSPs (800–1400 days), the LSP amplitudes averaged over 0.20 and ranged up to 0.40. For stars with larger LSP values than 1400 days, the LSP amplitudes averaged about 0.15, consistent with the values in Table 2.

There is still much to learn about the LSP phenomenon and, because it is a long-term phenomenon, AAVSO visual observations can be especially useful. Is the LSP constant in time? If it is a binary period, it should be. Does the LSP amplitude, and the LSP phase curve vary with time? This can provide information about variability in the amount and distribution of the dust around the companion. A decade or two ago, systematic radial velocities of red giants with LSPs were carried out. Comparing these with contemporaneous AAVSO photometry can ensure that the relative phase of the radial velocities and the photometry are consistent with the Soszyński *et al.* (2021) binary model. There is certainly a place for AAVSO observers in all of these projects.

The origin of LSP stars is still a bit of a mystery. Their short pulsation periods, and small pulsation amplitudes suggest that they are low-luminosity giants, which would have weak winds. In that case, however, how do they transfer enough matter to the companion for it to grow into a brown dwarf or low-mass star, and produce a significant LSP phenomenon?

## 5. Conclusions

We have presented and discussed the period and amplitude analysis of a random sample of 35 so-called LSP stars, in which the LSP variability is significantly larger than the pulsational variability. The pulsation period and amplitude increase with luminosity, as expected, and the LSP also increases with the luminosity as would be expected if the LSP were the orbital period of a close, dust-enshrouded companion. The LSP amplitude, which is related to the eclipse depth, is low in low-luminosity stars, perhaps because the effective size of the dust-enshrouded companion is smaller. The LSP amplitude is also low in high-luminosity stars, perhaps because the size of the stellar disc is large compared to the effective size of the companion. The LSP amplitude is largest, on average, for stars of moderate luminosity. The amplitude will also depend on the inclination of the companion's orbit to the line of sight, so at any $M\_K$, there will be a range of LSP amplitudes.

## 6. Acknowledgements

We thank the ASAS-SN project for making their data publicly available, and the creators and maintainers of the AAVSO VStar package for making it user-friendly and publicly available. We also thank the University of Toronto Work-Study Program and the Dunlap Institute for financial support. The Dunlap Institute is funded through an endowment created by the David Dunlap Family and the University of Toronto.

## References

Benn, D. 2013, VStar data analysis software (https://www.aavso.org/vstar-overview).
Derekas, A., Kiss, L. L., Bedding, T. R., Kjeldsen, H., Lah, P., and Szabó, Gy. M. 2006, *Astrophys. J., Lett.*, **650**, L55.
Houk, N. 1963, *Astron. J.*, **68**, 253.
Jayasinghe, T. *et al.* 2018, *Mon. Not. Roy. Astron. Soc.*, **477**, 3145.
Jayasinghe, T. *et al.* 2019, *Mon. Not. Roy. Astron. Soc.*, **486**, 1907.
Kim, J.V.E., and Percy, J. R. 2022, *J. Amer. Assoc. Var. Star Obs.*, **50**, 178.
Kochanek, C. S., *et al.* 2017, *Publ. Astron. Soc. Pacific*, **129**, 104502.
Lebzelter, T., and Wood, P. R. 2005, *Astron. Astrophys.*, **441**, 1117.
O'Connell, D. J. K. 1933, *Bull. Harvard Coll. Obs.*, No. 893, 19.
Pawlak, M. 2023, *Astron. Astrophys.*, **669A**, 60.
Payne-Gaposchkin, C. 1954, *Ann. Harvard Coll. Obs.*, **113**, 189.
Percy, J. R. 2022, University of Toronto, TSpace Repository (https://tspace.library.utoronto.ca/handle/1807/124406).
Percy, J. R., and Deibert, E. 2016, *J. Amer. Assoc. Var. Star Obs.*, **44**, 94.
Percy, J. R., and Gupta, P. 2021, *J. Amer. Assoc. Var. Star Obs.*, **49**, 209.
Percy, J. R., Landis, H. J., and Milton, R. E. 1989, *Publ. Astron. Soc. Pacific*, **101**, 893.
Percy, J. R., Wilson, J. B., and Henry, G. W. 2001, *Publ. Astron. Soc. Pacific*, **113**, 983.
Percy, J. R., and Zhitkova, S. 2023, *J. Amer. Assoc. Var. Star Obs.*, **51** (app.aavso.org/jaavso/article/3906/).
Samus, N. N., Kazarovets, E. V., Durlevich, O. V., Kireeva, N. N., and Pastukhova, E. N. 2017, *Astron. Rep.*, **61**, 80.
Shappee, B. J., *et al.* 2014, *Astrophys. J.*, **788**, 48 (https://asas-sn.osu.edu/).
Soszyński, I., *et al.* 2021, *Astrophys. J., Lett.*, **911**, L22.
Udalski, A., Szymański, M. K., and Szymański, G. 2015, *Acta Astron.*, **65**, 1.
Wood, P. R. 2000, *Publ. Astron. Soc. Australia*, **17**, 18.
Wood, P. R., Olivier, A. E., and Kawaler, S. D. 2004, in *Variable Stars in the Local Group*, eds. D. W. Kurtz, K. R. Pollard, ASP Conf. Proc. 310, 322.